\documentclass[useAMS]{gGAF2e}
\graphicspath{{./fig/}{./png/}}
\usepackage{natbib}
\def\la{\mathrel{\mathchoice {\vcenter{\offinterlineskip\halign{\hfil
$\displaystyle##$\hfil\cr<\cr\sim\cr}}}
{\vcenter{\offinterlineskip\halign{\hfil$\textstyle##$\hfil\cr<\cr\sim\cr}}}
{\vcenter{\offinterlineskip\halign{\hfil$\scriptstyle##$\hfil\cr<\cr\sim\cr}}}
{\vcenter{\offinterlineskip\halign{\hfil$\scriptscriptstyle##$\hfil\cr<\cr\sim\cr}}}}}

\newcommand{\apj}{\itshape Astrophys.\ J.\ }
\newcommand{\apjl}{\itshape Astrophys.\ J.\ Lett.\ }
\newcommand{\apjs}{\itshape Astrophys.\ J.\ Suppl.\ }
\newcommand{\aap}{\itshape Astron.\ Astrophys.\ }
\newcommand{\aapr}{\itshape Astron.\ Astrophys.\ Rev.\ }
\newcommand{\an}{\itshape Astron.\ Nachr.\ }
\newcommand{\mnras}{\itshape Monthly Notices of the Roy.\ Astron.\ Soc.\ }
\newcommand{\pre}{\itshape Phys.\ Rev.\ E }
\newcommand{\prl}{\itshape Phys.\ Rev.\ Lett.\ }
\newcommand{\jfm}{\itshape J.\ Fluid Mech.\ }

\newcommand{\etat}{\eta_{\rm t}}

\newcommand{\urms}{u_{\rm rms}}

\newcommand{\Beq}{B_{\rm eq}}

\newcommand{\kef}{k_{\rm f}}

\newcommand{\Sh}{{\rm Sh}}
\newcommand{\Pm}{{\rm Pm}}
\newcommand{\Rey}{{\rm Re}}
\newcommand{\Rm}{{\rm Rm}}
\newcommand{\Pra}{{\rm Pr}}
\newcommand{\Ma}{{\rm Ma}}
\newcommand{\Ra}{{\rm Ra}}

\newcommand{\Co}{{\rm Co}}

\newcommand{\yyy}{\bm{\hat{y}}}
\newcommand{\OO}{\bm{\varOmega}}
\newcommand{\UU}{\bm{U}}
\newcommand{\ccdot}{{\bm\cdot}}
\newcommand{\DIV}{\bm{\nabla} {\bm\cdot} }
\newcommand{\pd}{\partial}
\newcommand{\meanv}[1]{\overline{\bm #1}}
\newcommand{\mean}[1]{\overline{#1}}
\def\onethird{{\textstyle{1\over3}}}
\def\onehalf{{\textstyle{1\over2}}}

\begin{document}
\doi{10.1080/03091929.2012.715158}
\issn{1029-0419} \issnp{0309-1929} \jvol{107} \jnum{1--2} \jyear{2013} \jmonth{March}
\setcounter{page}{244}
\markboth{P. J. K\"APYL\"A \itshape{et al.}}{Oscillatory large-scale 
dynamos from Cartesian convection simulations}
\title{Oscillatory large-scale dynamos from Cartesian convection simulations}
\author{P. J. K\"APYL\"A${\dag\ddag}$$^{\ast}$\thanks{$^\ast$Corresponding 
author. Email: petri.kapyla@helsinki.fi
\vspace{6pt}} M. J. MANTERE${\dag}$ 
and A. BRANDENBURG${\ddag}$\\\vspace{6pt}  
${\dag}$Physics Department, Gustaf H\"allstr\"omin katu 2a, FI-00014 
University of Helsinki, Finland\\
${\ddag}$NORDITA, Royal Institute of Technology and Stockholm University,
Roslagstullsbacken 23, SE-10691 Stockholm, Sweden; and
%
Department of Astronomy, Stockholm University, SE-10691 
Stockholm, Sweden\\
\vspace{6pt}\received{\today, $Revision: 1.110 $}}

\maketitle

\begin{abstract}
  We present results from compressible Cartesian convection simulations 
  with and without imposed shear.
  In the former case the dynamo is expected to be of $\alpha^2\varOmega$
  type which is generally expected to be relevant for the Sun, whereas the
  latter case refers to $\alpha^2$ dynamos which are more likely to occur
  in more rapidly rotating stars whose differential rotation is small.
  We perform a parameter study where the
  shear flow and the rotational influence are
  varied to probe the relative importance of both types of dynamos.
  Oscillatory solutions are preferred both in the kinematic and
    saturated regimes when the negative ratio of shear to rotation rates,
    $q\equiv -S/\varOmega$, is between 1.5 and 2, i.e., when shear and
    rotation are of comparable strengths.
  Other regions of oscillatory solutions are found with small
    values of $q$, i.e., when shear is weak in comparison to
    rotation, and in the regime of large negative $q$s, when shear
    is very strong in comparison to rotation.
  However, exceptions to these rules also appear so that for a
    given ratio of shear to rotation, solutions are non-oscillatory
    for small and large shear, but oscillatory in the intermediate
    range. Changing the boundary conditions from vertical field to perfect
    conductor ones changes the dynamo mode from oscillatory to
    quasi-steady. Furthermore, in
  many cases an oscillatory solution exists only in the kinematic regime
  whereas in the nonlinear stage the mean fields are stationary. However, the
  cases with rotation and no shear are always oscillatory in the parameter range
  studied here and the dynamo mode does not depend on the magnetic boundary
  conditions. The strengths of total and large-scale components of
  the magnetic field in the saturated state, however, are sensitive to
  the chosen boundary conditions.
\end{abstract}
\bigskip

\bigskip

\section{Introduction}
The solar magnetic cycle is commonly thought to be a manifestation of
an oscillatory large-scale dynamo operating within or just below the
convection zone \citep[e.g.][]{O03}. A possible origin of the solar
magnetic fields is the turbulent dynamo mechanism, where helical
small-scale fluid motions and large-scale shear sustain the magnetic
field \citep[e.g.][]{M78,KR80,RH04}. According to mean-field theory,
turbulent stratified convection together with global rotation of
the Sun lead to an $\alpha$ effect \citep{SKR66} and large-scale 
differential
rotation \citep[e.g.][]{R89}. Their combined effect constitutes the
$\alpha\varOmega$-dynamo which often yields oscillatory solutions
\citep[e.g.][]{P55,SK69}.

However, reproducing the solar cycle with direct
numerical simulations still remains challenging
\citep[e.g.][]{MT09,K11}. Early spherical
simulations were indeed able to achieve oscillatory large-scale fields
that propagate toward the poles \citep{G83,G85}.
Similar results have been confirmed by a number of recent simulations
in spherical shells \citep{Brown10,Brown11,Charb10,Charb11} as well as
in wedges of spherical shells \citep{KKBMT10}.
Recent simulations in wedges have also yielded equatorward migration
\citep{KMB12}.
Simpler Cartesian
models with rotating stratified convection were less successful as
only small-scale fields were seen \citep[e.g.][]{Nea92,Bea96}. Only
when a shear flow was added \citep{KKB08,HP09} or rapid enough rotation
was used \citep{JR00,RJ02,KKB09b}, large-scale fields were
obtained. Even in the cases with imposed shear, no oscillatory
solutions were seen although the necessary prerequisites, helical
turbulence and shear, were present.
However, these are aspects that depend critically on the boundary conditions.
Indeed, \cite{KKB09c} have presented mean-field calculations of associated
convection simulations that agree with each other not only qualitatively
in that both are non-oscillatory, but they also agree quantitatively as
far as their excitation condition is concerned.

Here we extend previous studies on large-scale dynamos due to 
turbulent convection in Cartesian
geometry \citep{KKB08,KKB09b} to cover a larger parameter
space and explore more thoroughly the effects of boundary conditions on the
solutions. We present runs with imposed shear and find that
oscillatory solutions can be found in a limited part of the parameter
range studied. We also report on rigidly rotating runs where
oscillatory $\alpha^2$-dynamos are observed.

\section{Model}

Our model setup is the same as that of \cite{KKB08,KKB09b}.
A rectangular portion of a star is represented by
a box situated at colatitude $\theta$. The dimensions of the
computational domain are $(L_x, L_y, L_z) =
(4,4,2)d$, where $d$ is the depth of the convectively unstable layer,
which is also used as the unit of length. The box is divided into
three layers: an upper cooling layer, a convectively unstable layer,
and a stable overshoot layer (see below). 
The following equations for compressible magnetohydrodynamics are 
solved:
\begin{equation}
\frac{\mathcal{D} \bm A}{\mathcal{D}t} = -S A_y \hat{\bm{x}} - (\bm{\nabla}\bm{U})^{\rm T}\bm{A}-\eta \mu_0{\bm J}, \label{equ:AA}
 \end{equation}
\begin{equation}
\frac{\mathcal{D} \ln \rho}{\mathcal{D}t} = -\DIV{\bm U},
 \end{equation}
\begin{equation}
 \frac{\mathcal{D} \bm U}{\mathcal{D}t} = -SU_x\bm{\hat{y}}-\frac{1}{\rho}{\bm \nabla}p + {\bm g} - 2\bm{\varOmega}_0 \times \bm{U} + \frac{1}{\rho} \bm{J} \times {\bm B} + \frac{1}{\rho} \bm{\nabla} \ccdot 2 \nu \rho \mbox{\boldmath ${\sf S}$}, \label{equ:UU}
 \end{equation}
\begin{equation}
 \frac{\mathcal{D} e}{\mathcal{D}t} = - \frac{p}{\rho}\DIV {\bm U} + \frac{1}{\rho} \bm{\nabla} \ccdot K \bm{\nabla}T + 2 \nu \mbox{\boldmath ${\sf S}$}^2 + \frac{\mu_0\eta}{\rho} \bm{J}^2 - \frac{e\!-\!e_0}{\tau(z)}, \label{equ:ene}
 \end{equation}
where $\mathcal{D}/\mathcal{D}t = \pd/\pd t + (\bm{U} + \meanv{U}_0)
\ccdot \bm{\nabla}$ is the advective derivative with respect to the total
(turbulent plus shear) flow, $\meanv{U}_0 = (0,Sx,0)$ is the imposed
large-scale shear flow, $\bm{A}$ is the magnetic vector potential,
$\bm{B} = \bm{\nabla} \times \bm{A}$ is the magnetic field,
$\bm{J} =\bm{\nabla} \times \bm{B}/\mu_0$ is the current density,
$\mu_0$ is the magnetic permeability, $\eta$ and $\nu$ are the magnetic diffusivity and
kinematic viscosity, respectively, $K$ is the heat conductivity,
$\rho$ is the density, $\bm{U}$ is the
velocity, $\bm{g} = -g\hat{\bm{z}}$ is the gravitational acceleration,
and $\bm{\varOmega}_0=\varOmega_0(-\sin \theta,0,\cos \theta)$ is the rotation
vector.
The fluid obeys an ideal gas law $p=\rho e (\gamma-1)$, where $p$
and $e$ are the pressure and internal energy, respectively, and
$\gamma = c_{\rm P}/c_{\rm V} = 5/3$ is the ratio of specific heats at
constant pressure and volume, respectively.
The specific internal energy per unit mass is related to the
temperature via $e=c_{\rm V} T$.
The traceless rate of strain tensor $\mbox{\boldmath ${\sf S}$}$ is given by
\begin{equation}
{\sf S}_{ij} = \onehalf (U_{i,j}+U_{j,i}) - \onethird \delta_{ij} \DIV \bm{U}.
\end{equation}
The last term of equation~(\ref{equ:ene}) describes cooling at the top of
the domain. Here, $\tau(z)$ is a cooling time with a profile
smoothly connecting the upper cooling layer and the convectively
unstable layer below, where $\tau^{-1}(z)\to0$.
Let us note that equation~(\ref{equ:AA}) is here written in the fully
advective gauge, but in practice, to avoid excessive buildup of gradient
contributions to $\bm{A}$ \citep{CHBM11}, it is solved just in the
shearing-advective gauge; see \cite{HB11} for details.

The positions of the bottom of the box, bottom and top of the
convectively unstable layer, and the top of the box are given
respectively by $(z_1, z_2, z_3, z_4) = (-0.85, 0, 1, 1.15)d$. Initially
the stratification is piecewise polytropic with polytropic indices
$(m_1, m_2, m_3) = (3, 1, 1)$, which leads to a convectively unstable
layer above a stable layer at the bottom of the domain and an
isothermal cooling layer at the top.
All simulations with rotation use $\theta=0$, corresponding to
the north pole.
Our initial stratification is given by the associated hydrostatic
equilibrium solution \citep{Bea96}, where velocity and magnetic
fields are perturbed with Gaussian noise of low amplitude.

\subsection{Nondimensional quantities and parameters}
\label{Nondimensional}

Dimensionless quantities are obtained by setting
\begin{eqnarray}
d = g = \rho_0 = c_{\rm P} = \mu_0 = 1\;,
\end{eqnarray}
where $\rho_0$ is the initial density at $z_2$. The units of length, time,
velocity, density, entropy, and magnetic field are
\begin{eqnarray}
&& [x] = d\;,\;\; [t] = \sqrt{d/g}\;,\;\; [U]=\sqrt{dg}\;,\;\;
[\rho]=\rho_0\;,\;\; [s]=c_{\rm P}\;,\;\; [B]=\sqrt{dg\rho
_0\mu_0}\;. 
\end{eqnarray}
The equipartition magnetic field is defined by 
\begin{equation}
\Beq \equiv \langle\mu_0\rho\bm{U}^2\rangle^{1/2},\label{equ:Beq}
\end{equation}
where angle brackets denote volume averaging.
We define the fluid and magnetic Prandtl numbers and the Rayleigh
number as
\begin{eqnarray}
\Pra=\frac{\nu}{\chi_0}\;,\;\; \Pm=\frac{\nu}{\eta}\;,\;\; \Ra=\frac{gd^4}{\nu \chi_0} \bigg(-\frac{1}{c_{\rm P}}\frac{{\rm d}s}{{\rm d}z
} \bigg)_0\;,
\end{eqnarray}
where $\chi_0 = K/(\rho_{\rm m} c_{\rm P})$ is the thermal
diffusivity, and $\rho_{\rm m}$ is the density in the middle of
the convectively unstable layer, $z_{\rm m}=z_3-z_2$.
The entropy gradient, measured at $z_{\rm m}$ in the non-convecting
hydrostatic state, is given by
\begin{eqnarray}
\bigg(-\frac{1}{c_{\rm P}}\frac{{\rm d}s}{{\rm d}z}\bigg)_0 = \frac{\nabla-\nabla_{\rm ad}}{H_{\rm P}}\;,
\end{eqnarray}
where $\nabla-\nabla_{\rm ad}$
is the superadiabatic temperature gradient with 
$\nabla_{\rm  ad} = 1-1/\gamma$, $\nabla = (\pd \ln T/\pd \ln
  p)_{z_{\rm m}}$, where $H_{\rm P}$ is the pressure scale height.
The degree of stratification is determined by the parameter 
$\xi_0 =(\gamma-1) e_0/gd$, which is the pressure scale height at
the top of the domain normalized by the depth of the unstable layer.
We use $\xi_0 =0.3$ in all cases,
which results in a density contrast of about 23.
We define the fluid and magnetic Reynolds numbers via
\begin{eqnarray}
\Rey = \frac{\urms}{\nu \kef}\;,\quad
{\rm Rm} = \frac{\urms}{\eta \kef}\;,
\end{eqnarray}
where $\kef = 2\pi/d$ is assumed as a reasonable estimate
for the wavenumber of the energy-carrying eddies.
Note that, according to this definition, $\Rm$ is by a factor $2\pi$
smaller than the usually adopted one based on $d$ instead of $\kef$.
The amounts of shear and rotation are quantified by
\begin{eqnarray}
{\rm Sh} = \frac{S}{\urms \kef}\;,\quad
{\rm Co} = \frac{2\,\varOmega_0}{\urms \kef}\;. \label{equ:ShCo}
\end{eqnarray}
The denominators in equation~(\ref{equ:ShCo}) give an estimate of the 
inverse convective turnover time.
We also use the value of the relative shear rate
\begin{equation}
q=-S/\varOmega_0=-2\frac{{\rm Sh}}{{\rm Co}},
\end{equation}
which is often used in the context of disk systems, for which the
local angular velocity varies like $\varOmega(r) \propto r^{-q}$
with $q=1.5$.
For $q\geq2$ the flow is Rayleigh unstable.
Note that for $q=2$, we have $2\varOmega_0+S=0$,
so the effect of rotation and shear,
$SU_x\yyy+2\OO\times\UU=\left(-2\varOmega_0 U_y,(2\varOmega_0+S)U_x,0\right)$,
reduces to $(\tilde{S}U_y,0,0)$.
Here we have introduced the quantity $\tilde{S}=-2\varOmega_0$ to highlight
the similarity to plane shear flow ($\varOmega_0=0$), in which the effect of
shear is given by $(0,SU_x,0)$.
This analogy between plane shear flow and marginally Rayleigh-stable
flows was noted by \cite{BHS96} and will also play a role in our
considerations below.

\subsection{Boundary conditions}

Stress-free boundary conditions are used for the velocity,
\begin{equation}
U_{x,z} = U_{y,z} = U_z = 0,
\end{equation}
and either vertical field or perfect conductor conditions for the
magnetic field, i.e.\ 
\begin{eqnarray}
B_x = B_y &=& 0 \;\; {\rm (vertical\;field)}, \\
B_{x,z} = B_{y,z} = B_z &=& 0 \;\; {\rm (perfect\;conductor)},
\end{eqnarray}
respectively.
We may think of them as open and closed boundaries, respectively,
because they either do or do not permit a magnetic helicity flux.
In the $y$ and $x$ directions we use periodic and shearing-periodic
boundary conditions, respectively.
In the runs with shear and rotation we use vertical field
conditions at the top and perfect conductor conditions at the bottom,
unless stated otherwise.
The simulations have been made with the {\sc Pencil Code}%
\footnote{\texttt{http://pencil-code.googlecode.com/}}
using sixth-order explicit finite differences in space and a third
order accurate time stepping method.

\section{Results}
We perform three sets of simulations with our standard setup
with shear and rotation
(Sets~A, B, and C) and a few exploratory
runs with only rotation (Set~D); see Table~\ref{tab:runs}. 
In the 
  less extensive Sets~E and F (see Table~\ref{tab:runs2}) we explore
  the parameter regime in the vicinity of one of the oscillatory
  models Run~A2 (Set~E), and the effect of changing boundary conditions
  on the solution (Set~F).
In Sets~A and B we keep the shear rate $S$
constant and vary the rotation rate $\varOmega_0$. In Set~A we use
$S=-0.05\sqrt{g/d}$ and in Set~B we have $S=-0.1\sqrt{g/d}$. In
  Set~C the rotation rate $\varOmega_0=0.1\sqrt{g/d}$ is fixed and the
  shear rate $S$ is varied. Our hydrodynamical progenitors of the runs
in Set~A were taken from \cite{KBKSN10} and those of the runs in Set~B were
obtained by doubling both $S$ and $\varOmega_0$. In terms of $q$, we
explore the range $-10\ldots1.99$.
We take Runs~A9 and A1 from \cite{KMH11} as the hydrodynamical progenitors
for our runs in Set~D with only rotation.

The case $S\neq0$ and $\varOmega_0=0$ corresponds to $q\to\pm\infty$ and is
a special case in which a `vorticity dynamo' \citep[e.g.][]{Eea03,KMB09}
is excited for the values of shear chosen here.
In this part of the parameter range we use data from
\cite{KKB08} with $\varOmega_0=0$ and $S\neq 0$.
Values of $q$ near zero refer to runs with rapid and nearly 
rigid rotation, whereas large values of $|q|$ are associated with 
strong shear and slow rotation.
For $q\geq2$ the flow is
Rayleigh unstable and thus we limit our study to values $q\leq1.99$.
However, in Set~C we find large-scale vorticity generation for
  $q=1.99$, leading eventually to supersonic velocities and numerical
  instability.
  In view of the analogy between plane shear flow and marginally
  Rayleigh-stable flows (Sect.~\ref{Nondimensional}), this large-scale vorticity
  generation might be related to the aforementioned vorticity dynamo.
  Therefore we reduce the
  highest value of $q$ to $1.95$ in Set~C. In many runs with $q>0$ the
  rms-velocity increases in the saturated regime which is likely due
  to the magnetorotational instability. This is particularly relevant
  in cases where the shear is strong, i.e.\ runs in Sets~B and C.

\begin{table}
  \tbl{Summary of the runs. $\Ma_{\rm kin}$ and $\Ma$ are the volume 
    averaged rms-velocities from the kinematic and saturated states, 
    respectively.
    In Sets~A and B we use $\Pra=1$, $\Ra=10^6$, 
    $\Pm=1$, and grid resolution $128^3$. Perfect conductor (vertical 
    field) conditions for the magnetic field at the lower (upper) boundary 
    are used.
    In Set~D, $\Pm$ varies, while $\Pra=0.24$, $\Ra=2\cdot10^6$, $\Sh=0$, and 
    grid resolution $256\times128^2$. The boundary conditions in Set~D 
    are listed in the rightmost column of the table. 
    Oscillatory (osc) and stationary (sta) modes of the dynamo are 
    denoted in the second column from right.
    Oscillatory decay is marked by ``osc/$-$'' with the comment ``no dynamo''.
    For both types of oscillatory solutions, the $q$ values are indicated
    in italics.
    Question marks indicate
    that only very few sign changes are covered by the time series or 
    irregular reversals are seen.
    Here, 
    $\tilde{B}_{\rm rms}={B}_{\rm rms}/B_{\rm eq}$, where $B_{\rm rms}$ is the 
    total rms magnetic field, and
    $\tilde{\mean B}_i={\langle\mean{B}_i^2}\rangle^{1/2}/B_{\rm eq}$.}
{\begin{tabular}{@{}lcccrrccccll}\toprule
   Run & $\Ma_{\rm kin}$ & $\Ma$ & $\Rm$ & $q\;\;$  & $\!\Co\;$ &   $\Sh$ & $\tilde{B}_{\rm rms}$ & $\tilde{\mean B}_x$ & $\tilde{\mean B}_y$ & Mode & Comment \\
\colrule
   A1  & 0.036 & 0.031 &   25 &{\em1.99}& 0.26 & $-0.25$ & 1.50 & 0.07 & 1.38 & osc/osc? & \\ 
   A2  & 0.034 & 0.029 &   23 &{\em1.75}& 0.31 & $-0.27$ & 1.50 & 0.08 & 1.27 & osc/osc \\ 
   A3  & 0.032 & 0.028 &   22 &{\em1.50}& 0.38 & $-0.28$ & 1.58 & 0.10 & 1.29 & osc/osc \\ 
   A4  & 0.030 & 0.027 &   22  &  1.25 &  0.46 & $-0.29$ & 1.41 & 0.10 & 1.12 & sta/osc? \\ 
   A5  & 0.032 & 0.028 &   22  &  1.00 &  0.57 & $-0.28$ & 2.65 & 0.22 & 2.46 & sta/sta \\ 
   A6  & 0.029 & 0.029 &   23  &  0.75 &  0.73 & $-0.27$ & 3.34 & 0.29 & 3.16 & sta/sta \\ 
   A7  & 0.031 & 0.030 &   24  &  0.50 &  1.07 & $-0.27$ & 3.92 & 0.37 & 3.73 & sta/sta \\ 
   A8  & 0.028 & 0.029 &   23  &  0.25 &  2.17 & $-0.27$ & 4.34 & 0.47 & 4.08 & sta/sta \\ 
   A9  & 0.017 & 0.026 &   20 &{\em0.10}& 6.19 & $-0.31$ & 5.70 & 0.61 & 5.33 & osc/sta \\ 
   A10 & 0.010 & 0.023 &   18 &{\em0.05}& 13.7 & $-0.34$ & 7.07 & 0.75 & 6.53 & osc/sta \\ 
   A11 & 0.011 & 0.011 &    8&$-${\em0.05}& $-30.2$ & $-0.75$ & $-$ & $-$ & $-$ & osc/$-$ & not run to saturation \\ 
   A12 & 0.017 & 0.016 &   13&$-${\em0.10}& $-9.87$ & $-0.49$ & $-$ & $-$ & $-$ & osc/$-$ & not run to saturation \\ 
   A13 & 0.023 & 0.056 &   45&$-${\em0.25}& $-1.13$ & $-0.14$ & 2.21 & 0.36 & 1.81 & osc?/sta \\ 
   A14 & 0.026 & 0.062 &   49  & $-0.50$ &  $-0.52$ & $-0.13$ & 2.10 & 0.22 & 1.85 & sta/sta \\ 
   A15 & 0.028 & 0.027 &   21&$-${\em1.0}\;\,&  $-0.59$ & $-0.30$ & $-$ & $-$ & $-$ & sta/$-$ & no dynamo \\ 
   A16 & 0.030 & 0.028 &   23&$-${\em2.5}\;\,&  $-0.22$ & $-0.28$ & $-$ & $-$ & $-$ & osc/$-$ & no dynamo \\ 
   A17 & 0.032 & 0.030 &   24&$-${\em5.0}\;\,&  $-0.11$ & $-0.26$ & $-$ & $-$ & $-$ & osc/$-$ & no dynamo \\ 
   A18 & 0.037 & 0.036 &   28&$-${\em10.0}\;\,& $-0.04$ & $-0.22$ & $-$ & $-$ & $-$ & osc/$-$ & no dynamo \\ 
\hline
   B1  & 0.044 & 0.090 &   72 &{\em1.99}& 0.18 & $-0.18$ & 1.15 & 0.09 & 1.00 & osc/sta \\ 
   B2  & 0.036 & 0.040 &   32 &{\em1.75}& 0.46 & $-0.40$ & 2.70 & 0.13 & 2.54 & osc/osc? \\ 
   B3  & 0.031 & 0.039 &   31  &  1.50 &  0.54 & $-0.40$ & 3.18 & 0.18 & 2.99 & sta/sta \\ 
   B4  & 0.030 & 0.040 &   31  &  1.25 &  0.64 & $-0.40$ & 3.22 & 0.20 & 3.03 & sta/sta \\ 
   B5  & 0.032 & 0.042 &   33  &  1.00 &  0.77 & $-0.38$ & 2.94 & 0.20 & 2.74 & sta/sta \\ 
   B6  & 0.027 & 0.034 &   27  &  0.75 &  1.25 & $-0.47$ & 3.79 & 0.34 & 3.57 & sta/sta \\ 
   B7  & 0.025 & 0.038 &   31  &  0.50 &  1.66 & $-0.42$ & 3.63 & 0.29 & 3.38 & sta/sta \\ 
   B8  & 0.019 & 0.036 &   29 &{\em0.25}& 3.50 & $-0.44$ & 4.00 & 0.33 & 3.68 & osc/sta \\ 
   B9  & 0.011 & 0.029 &   23  &  0.10 &  11.0 & $-0.55$ & 4.51 & 0.41 & 4.00 & sta/sta \\ 
   B10 & $-$ & $-$ &   $-$  & 0.05 &  $-$ & $-$ & $-$ & $-$ & $-$ & $-$ & no convection \\ 
   B11 & $-$ & $-$ &   $-$  & $-0.05$ &  $-$ & $-$ & $-$ & $-$ & $-$ & $-$ & no convection \\ 
   B12 & 0.011 & 0.011 &    8&$-${\em0.10}&$-30.2$ & $-1.51$ & $-$ & $-$ & $-$ & osc/$-$ & marginal dynamo \\ 
   B13 & 0.020 & 0.018 &   15&$-${\em0.25}&$-6.91$ & $-0.86$ & $-$ & $-$ & $-$ & osc?/$-$ & not run to saturation \\ 
   B14 & 0.024 & 0.086 &   69 &$-${\em0.50}& $-0.74$ & $-0.18$ & 1.35 & 0.19 & 1.01 & osc?/sta & \\ 
   B15 & 0.027 & 0.094 &   75 &$-${\em1.0}\;\,& $-0.34$ & $-0.17$ & 1.16 & 0.10 & 0.95 & osc?/sta & \\ 
   B16 & 0.032 & 0.030 &   24 &$-${\em2.5}\;\,& $-0.43$ & $-0.54$ & $-$ & $-$ & $-$ & osc/$-$ & marginal dynamo \\ 
   B17 & 0.037 & 0.032 &   26 &$-${\em5.0}\;\,& $-0.20$ & $-0.50$ & 1.44 & 0.06 & 1.20 & osc/sta & \\ 
   B18 & 0.042 & 0.035 &   28 &$-${\em10.0}\;\,&$-0.09$ & $-0.46$ & 1.61 & 0.08 & 1.35 & osc/sta & \\ 
\hline
   C1  & 0.055 & 0.165 &  131 &{\em1.95}&$0.19$ & $-0.19$ & 1.06 & 0.10 & 0.70 & sta/osc & \\ 
   C2  & 0.034 & 0.060 &   58 & 1.75 & 0.53 & $-0.46$ & 2.32 & 0.15 & 1.99 & sta/sta & \\ 
   C3  & 0.031 & 0.050 &   50 & 1.50 & 0.64 & $-0.48$ & 2.63 & 0.17 & 2.36 & sta/sta & \\ 
   C4  & 0.029 & 0.042 &   33 & 1.25 & 0.76 & $-0.47$ & 3.22 & 0.20 & 3.00 & sta/sta & \\ 
   C5  & 0.029 & 0.039 &   31 & 1.00 & 0.83 & $-0.41$ & 3.48 & 0.23 & 3.28 & sta/sta & \\ 
   C6  & 0.029 & 0.034 &   27 & 0.75 & 0.94 & $-0.35$ & 3.91 & 0.27 & 3.73 & sta/sta & \\ 
   C7  & 0.029 & 0.030 &   24 & 0.50 & 1.06 & $-0.27$ & 4.14 & 0.36 & 3.97 & sta/sta & \\ 
   C8  & 0.028 & 0.024 &   19 & 0.25 & 1.34 & $-0.17$ & 3.35 & 0.54 & 3.11 & sta/sta & \\ 
   C9  & 0.027 & 0.020 &   16 & 0.10 & 1.58 & $-0.08$ & 2.18 & 0.61 & 1.85 & sta/sta & \\ 
   C10 & 0.026 & $-$ &   21 & 0.05 & 1.20 & $-0.03$ & $-$ & $-$ & $-$ & sta/$-$ & not run to saturation \\ 
   C11 & 0.026 & $-$ &   21 & $-0.05$ & 1.21 & 0.03 & $-$ & $-$ & $-$ & sta/$-$ & no dynamo \\ 
   C12 & 0.027 & $-$ &   21 & $-0.10$ & 1.18 & 0.06 & $-$ & $-$ & $-$ & sta/$-$ & no dynamo \\ 
   C13 & 0.027 & $-$ &   21 & $-0.25$ & 1.18 & 0.15 & $-$ & $-$ & $-$ & sta/$-$ & not run to saturation \\ 
   C14 & 0.027 & $-$ &   22 & $-0.50$ & 1.16 & 0.29 & $-$ & $-$ & $-$ & sta/$-$ & not run to saturation \\ 
   C15 & 0.028 & $-$ &   23 &$-1.0$\;\,& 1.12 & 0.56 & $-$ & $-$ & $-$ & sta/$-$ & not run to saturation \\ 
   C16 & 0.033 & $-$ &   27 &$-2.0$\;\,& 0.95 & 0.95 & $-$ & $-$ & $-$ & sta/$-$ & not run to saturation \\ 
   C17 & 0.061 & $-$ &   48 &$-${\em5.0}\;\,& 0.52 & 1.31 & $-$ & $-$ & $-$ & osc/$-$ & marginal dynamo \\ 
\hline
   D1  & 0.081 & 0.021 &   66  & {\em0}$\quad$&  4.60 &     0 & 1.18 & 0.30 & 0.31 & osc/osc & $\Pm=2$, pc/vf \\ 
   D1b & $-$ & 0.025 &   39  & {\em0}$\quad$&  3.85 &     0 & 0.36 & 0.09 & 0.09 & osc/osc & $\Pm=1$, pc/vf \\ 
   D1c & $-$ & 0.032 &   26  & {\em0}$\quad$&  2.96 &     0 & 0 & 0 & 0 & osc/$-$ & \\ 
   D1d & 0.083 & 0.021 &   66  & {\em0}$\quad$&  4.60 &     0 & 1.18 & 0.32 & 0.29 & osc/osc & $\Pm=2$, vf/vf \\ 
   D1e & 0.083 & 0.023 &   72  & {\em0}$\quad$&  4.22 &     0 & 0.54 & 0.10 & 0.09 & osc/osc & $\Pm=2$, pc/pc \\ 
   D2  & 0.057 & 0.018 &   58  & {\em0}$\quad$&  17.5 &     0 & 0.64 & 0.10 & 0.10 & osc/osc & $\Pm=2$, pc/vf \\ 
   \botrule
  \end{tabular}}
\label{tab:runs}
\end{table}

\begin{table}
  \tbl{Summary of the additional runs.}
{\begin{tabular}{@{}lcccrrccccll}\toprule
   Run & $\Ma_{\rm kin}$ & $\Ma$ & $\Rm$ & $q\;\;$  & $\!\Co\;$ &   $\Sh$ & $\tilde{B}_{\rm rms}$ & $\tilde{\mean B}_x$ & $\tilde{\mean B}_y$ & Mode & Comment \\
\colrule
   E1  & 0.031 & 0.022 &   18 &   1.75 & 0.14 & $-0.12$ & 1.18 & 0.11 & 0.90 & sta/sta & \\ 
   E2  & 0.033 & 0.027 &   21 &{\em1.75}& 0.24 & $-0.21$ & 1.30 & 0.08 & 1.06 & osc/osc \\ 
   E3  & 0.034 & 0.029 &   23 &{\em1.75}& 0.31 & $-0.27$ & 1.50 & 0.08 & 1.27 & osc/osc & same as Run~A2 \\ 
   E4  & 0.037 & 0.031 &   25 &{\em1.75}& 0.41 & $-0.36$ & 2.09 & 0.10 & 1.87 & osc/osc \\ 
   E5  & 0.037 & 0.053 &   42 &   1.75 & 0.51 & $-0.45$ & 2.72 & 0.15 & 2.47 & sta/sta & \\ 
\hline
   F1  & 0.034 & 0.029 &   23 &{\em1.75}& 0.31 & $-0.27$ & 1.50 & 0.08 & 1.27 & osc/osc & pc/vf, same as Run~A2 \\ 
   F2  & 0.034 & 0.029 &   23 &{\em1.75}& 0.31 & $-0.27$ & 0.82 & 0.06 & 1.33 & osc/osc & vf/vf \\ 
   F3  & 0.034 & 0.029 &   23 &   1.75 & 0.31 & $-0.27$ & 1.51 & 0.10 & 1.33 & sta/sta & pc/pc \\ 
   \botrule
  \end{tabular}}
\label{tab:runs2}
\end{table}

In the following we discuss first the case with shear and study
the behavior of solutions for a range of values of $q$.
We refer to these solutions as $\alpha\varOmega$ (or $\alpha$--shear) dynamos.
We study separately the case without shear and refer to
such solutions as $\alpha^2$ dynamos,
which are generally also known as $\alpha^2\varOmega$ dynamos.
We focus here specifically on the case of oscillatory solutions.

\subsection{$\alpha^2\varOmega$ dynamos}
In our previous studies of convection-driven large-scale dynamos with
shear \citep{KKB08,KKB09b,KKB10}, only non-oscillatory solutions were
obtained. The large-scale field often had opposite signs in the
convectively unstable and stable layers, although
solutions with a single sign were obtained at low magnetic
Reynolds numbers \citep[see Fig.~5 of][]{KKB10} and in cases where
$\varOmega_0=0$ \citep[see Fig.~7 of][]{KKB08}. 
Runs with sinusoidal shear and rotation have also been reported to
show non-oscillatory large-scale fields \citep{HP09,KKB10c}.  
In this paper, we extend the parameter ranges of our previous
  studies in search of oscillatory solutions.
In this section we describe
  the results from different sets of runs
  individually and summarize the results for the different dynamo modes
  in Sect.~\ref{Summary31}.

\subsubsection{$S={\rm const}$, $\varOmega_0$ varies (Sets A and B)}

\begin{figure}
\begin{center}
\begin{minipage}{170mm}
\subfigure[]{
\resizebox*{8.5cm}{!}{\includegraphics{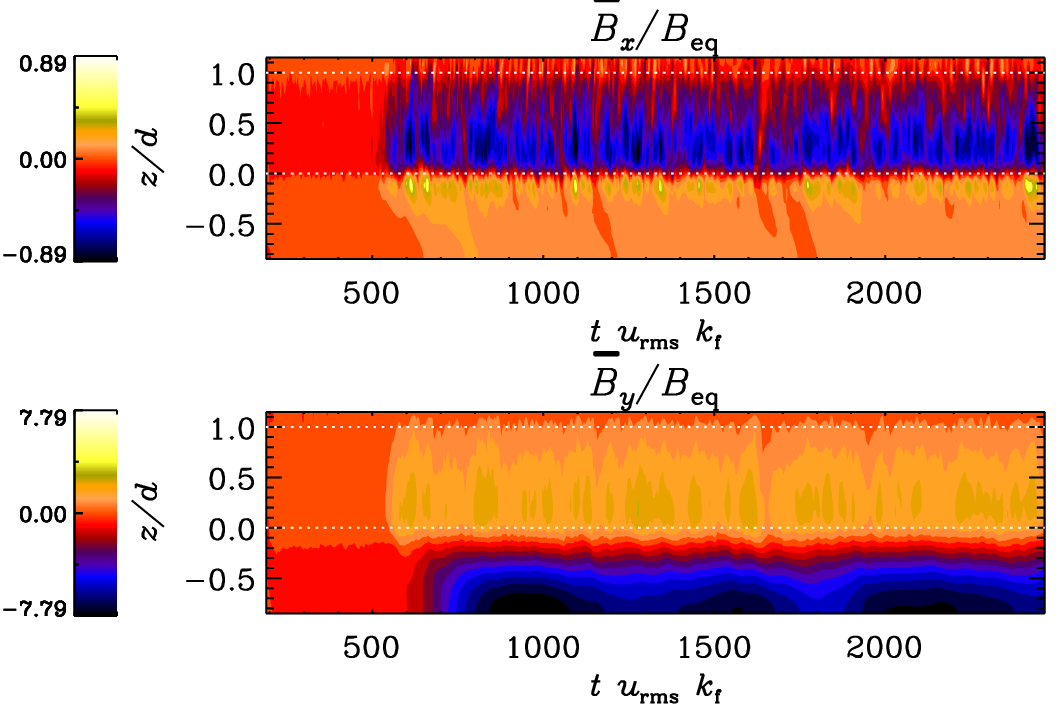}}}%
\subfigure[]{
\resizebox*{8.5cm}{!}{\includegraphics{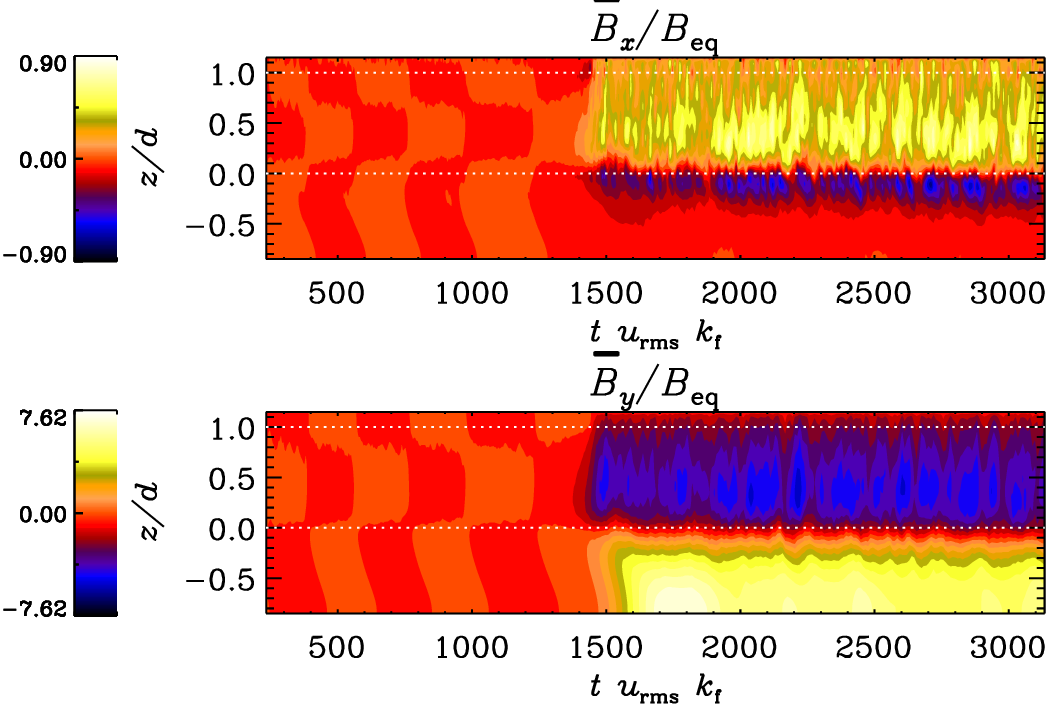}}}%
\caption{Horizontally averaged horizontal components of the magnetic
  field from non-oscillatory Run~A6 (a) and initially oscillatory but
  ultimately stationary Run~B8 (b) $\alpha$-shear dynamos.}%
\label{fig:butter1}
\end{minipage}
\end{center}
\end{figure}

As is evident from Table~\ref{tab:runs}, many of the runs in
Sets A and B are non-oscillatory; see Figure~\ref{fig:butter1}(a) for a
representative result from Run~A6.
For Set~A the mean magnetic field shows reversals in the range
  $1.5\la q<2$, while the increased $\varOmega_0$ in Set~B pushes the
  oscillatory regime somewhat towards higher $q$s.
The oscillations are particularly clear in the kinematic regime in
Runs~A1--A3. Cycles in the non-linear regime are more irregular and 
in one case 
the dynamo mode changes to a non-oscillatory mode (Run~B1).
We find that in the parameter regime
explored here, non-oscillatory solutions are excited in the range
$0.2\la q<1.5$.  
For $-10 \leq q \la 0.25$, however, another oscillatory regime is
found. These runs tend to show oscillations in the kinematic stage but
often switch to a stationary mode in the non-linear regime; see the
time evolution of the horizontally averaged magnetic field components
of Run~B8 in Figure~\ref{fig:butter1}(b).
  This might just be a manifestation of the fact that the excitation
  conditions for different dynamo modes do not necessarily reflect
  the stable dynamo mode in the nonlinear regime \citep{BKMMT89}.
This behaviour also explains the lack of oscillatory dynamos
in our previous works where we always used $q=1$.
This further illustrates the importance of comprehensive parameter
studies instead of individual numerical experiments \citep{KBKSN10}.

\begin{figure}
\begin{center}
\begin{minipage}{170mm}
\subfigure[]{
\resizebox*{17cm}{!}{\includegraphics{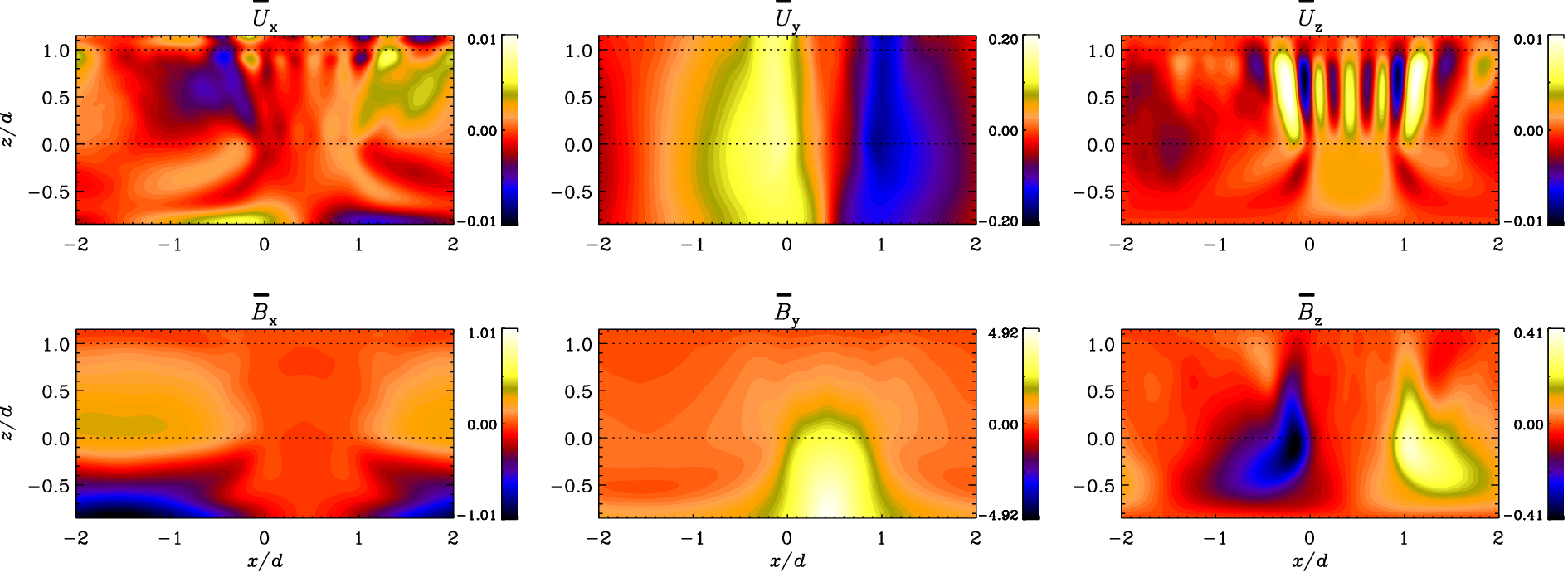}}}%
\caption{Mean velocities (upper row) and magnetic fields (lower row)
  averaged over the $y$-direction and time from Run~B14. The
  velocities are shown in units of $\sqrt{dg}$ and the magnetic fields
  in units of volume averaged equipartition field $B_{\rm eq}$.}%
\label{fig:pub}
\end{minipage}
\end{center}
\end{figure}

Convection is suppressed especially near $q=0$ due to the
rapid rotation, decreasing the Reynolds number and thus also leading to the
absence of dynamo action in this regime.
Many runs in the $q<0$ regime, especially in Set~A, are either
subcritical or show very slow growth of the magnetic field and were 
not run to
saturation. 
This lack of dynamo action may seem surprising because in the $q<0$
regime the contributions to the $\alpha$ effect due to shear and
rotation have the same sign \citep{KKB09b}. On the other hand, the
magnetorotational instability can be excited for $q>0$ which may
explain the more favourable dynamo excitation in that regime.
However, we find that if a saturated dynamo is present in this regime,
also the turbulence is enhanced (see Runs~A13, A14, B14, and B15).
This is associated with the generation of additional large-scale flows
that depend on $x$; see Figure~\ref{fig:pub}. The large-scale
magnetic fields are generally also $x$-dependent. Such modes are not 
visible in the kinematic stages of the runs.

\subsubsection{$S$ varies, $\varOmega_0={\rm const}$ (Set C)}
In Set~C, only Run~C17 with the strongest shear shows
  oscillations in the kinematic regime. The large-scale field in
  Run~C1 shows oscillations in the nonlinear stage. 
We note that this is the only stable--to--oscillatory transition
in the whole series of models explored.
In the kinematic stage of this run, we observe an rms-velocity enhanced
by almost a factor of two in comparison to Run~C2. The only difference
between these runs is a 10 percent smaller value of $S$ for Run~C2. 
The enhancement is likely to be due to the large-scale flows
generated by the vorticity dynamo.  
Furthermore, as the magnetic field
grows, the rms-velocity increases by another factor of three, which may
be attributed to the magnetorotational instability. Similarly as in
Sets~A and B, the dynamo is harder to excite for $q<0$ and in many
cases the growth rate of the magnetic field is low, which is the reason 
why some of the runs were not continued up to saturation.
  
\subsubsection{Sets E and F}
It appears that small changes of $\Co$ and $\Sh$ are enough to
  change the dynamo mode, e.g.\ compare Runs~B2 and C2 with
  $q=1.75$. In Set~E we vary $\Co$ and $\Sh$, taking an oscillatory
  Run~A2 as our basis (see Table~\ref{tab:runs2}). We find that
  oscillatory solutions are found in the intermediate range
  $0.24<\Co<0.41$, corresponding to $0.21<-\Sh<0.36$, whereas for lower and higher
  values of $\Co$ and $\Sh$ quasi-steady solutions appear.

  In Set~F we vary the magnetic boundary conditions of Run~A2 and find
  that for perfect conductor boundaries at the top and bottom of the
  domain, the solution is non-oscillatory. For a vertical field
  condition on both boundaries oscillatory solution is found.

\subsubsection{Cycle frequency and phase diagrams}
\begin{figure}
\begin{center}
\begin{minipage}{170mm}
\subfigure[]{
\resizebox*{8.5cm}{!}{\includegraphics{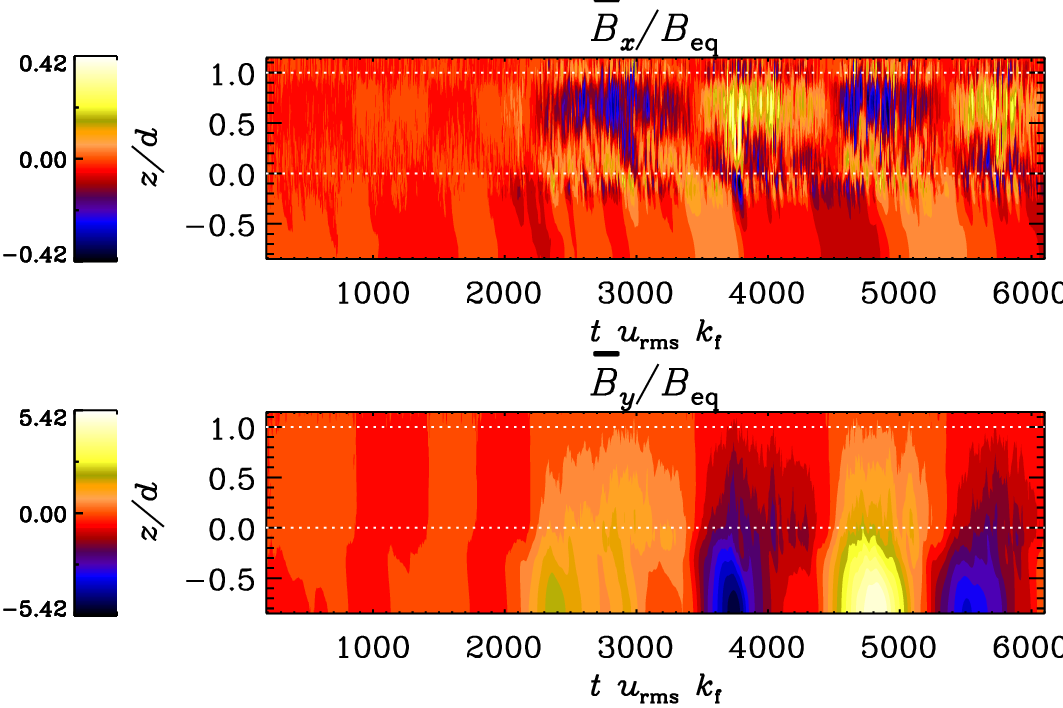}}}%
\subfigure[]{
\resizebox*{8.5cm}{!}{\includegraphics{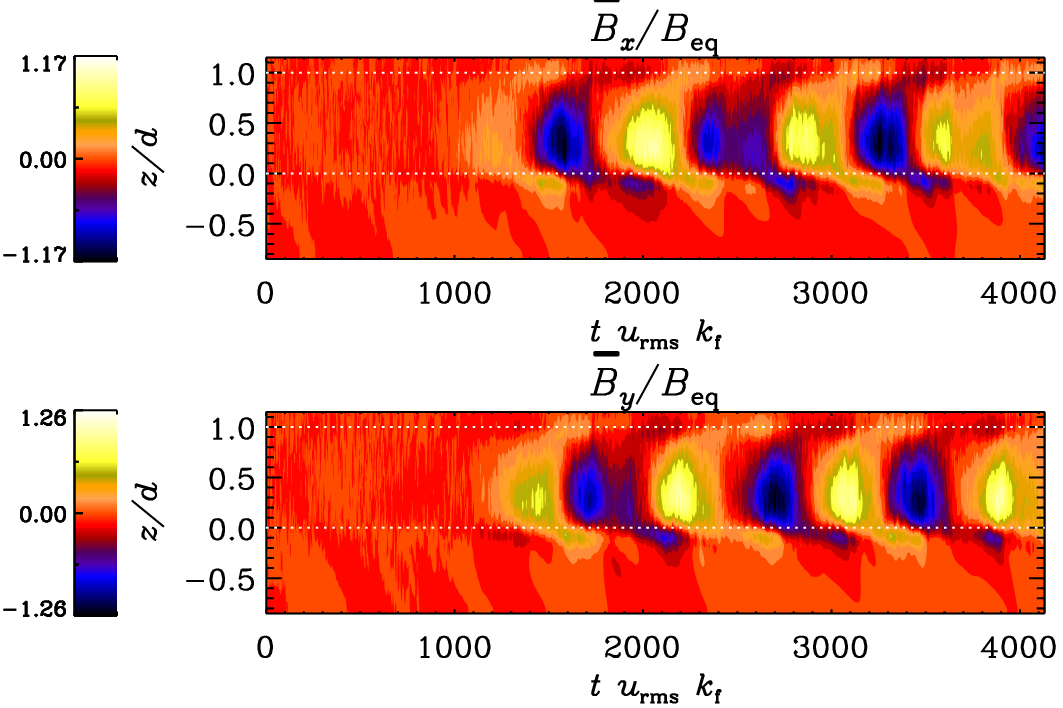}}}%
\caption{Same as Fig.~\ref{fig:butter1} but for oscillatory
  $\alpha$-shear dynamo Run~A2 (a) and $\alpha^2$ dynamo Run~D1 (b).}%
\label{fig:butter2}
\end{minipage}
\end{center}
\end{figure}

Even in the cases with the clearest oscillatory solutions, e.g.\ Run~A2 in
Fig.~\ref{fig:butter2}(a), the period of the oscillation varies from
cycle to cycle. Furthermore, the cycle period is of the order of
$10^3$ convective turnover times in this run.
Such a long cycle period limits the
duration of the simulation to only a few cycles.

The cycle frequency of a saturated $\alpha$--shear dynamo under the
assumption of homogeneity is given by 
\begin{equation}
\omega_{\rm cyc}=\eta_{\rm T} k_{\rm m}^2,
\label{omcyc}
\end{equation}
where $\eta_{\rm T}=\eta_{\rm t}+\eta$ is the total magnetic
diffusivity, $\eta_{\rm t}$ is the turbulent diffusivity, and $k_{\rm
  m}$ is the wavenumber of the dominant mode of the magnetic field
\citep{BB02}.
Empirically, a similar law
($\omega_{\rm cyc}/\eta_{\rm T} k_{\rm m}^2=1.6$...2.3)
was also found for oscillatory $\alpha^2$ dynamos
with nonuniform helicity distribution \citep{BCC09}.
Since equation~(\ref{omcyc}) is also valid in the non-linear regime, the
quenching of $\eta_{\rm t}$ as a function of $\Rm$ and $\mean{\bm B}$
can be estimated \citep{KB09}.
Although equation~(\ref{omcyc}) is expected to be inaccurate in the present
case where the value of $k_{\rm m}$ is uncertain, the cycle
frequency is likely regulated by the value of $\etat$. This suggests
that the turbulent diffusivity is quenched by a factor of roughly two
to three in Run~A2, compared to the kinematic stage of the same run.
This is entirely consistent with independent measurements of $\etat$
using the quasi-kinematic test-field method for quenched $\alpha^2$
dynamos \citep{BRRS08}.

\begin{figure}
\begin{center}
\begin{minipage}{170mm}
\subfigure[]{
\resizebox*{8.5cm}{!}{\includegraphics{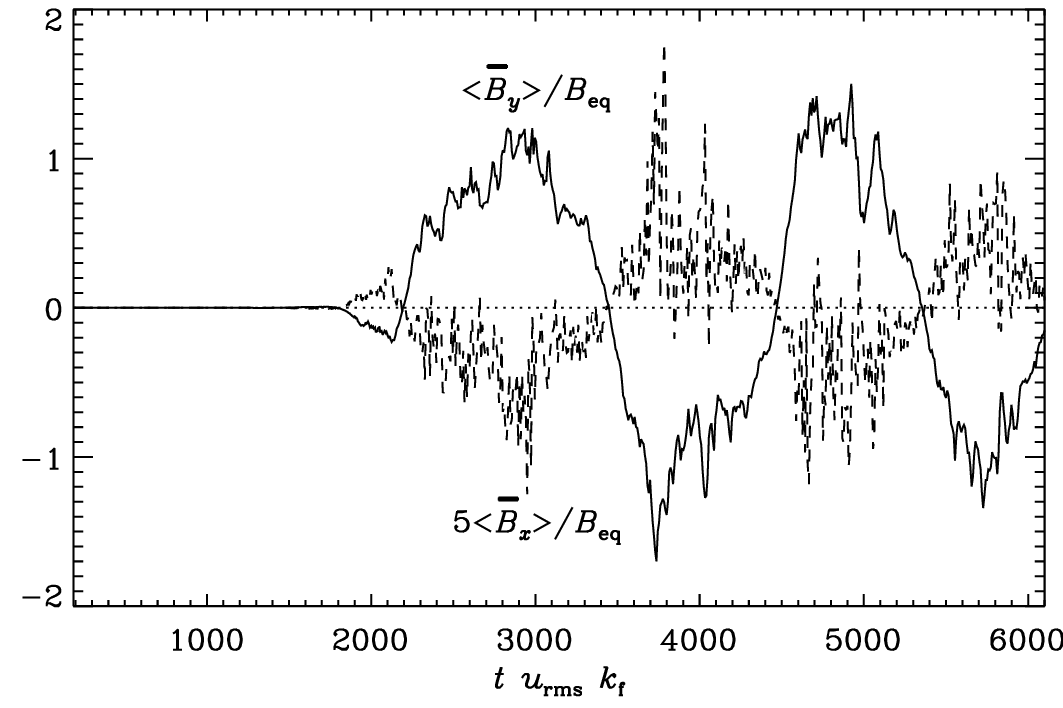}}}%
\subfigure[]{
\resizebox*{8.5cm}{!}{\includegraphics{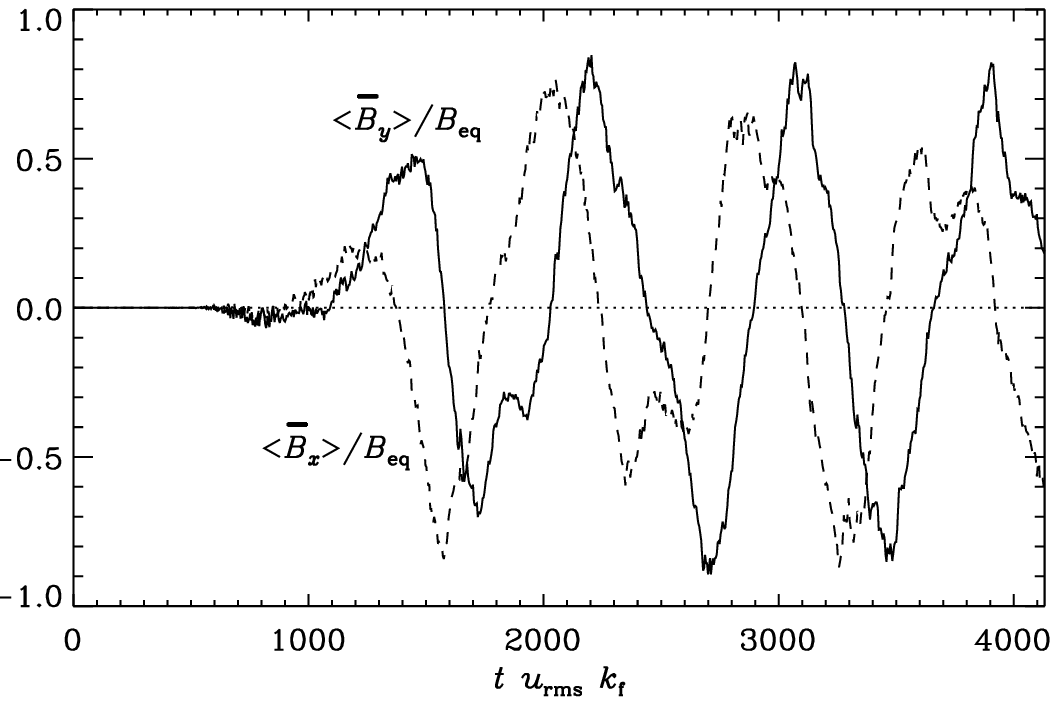}}}%
\caption{Phase diagrams for the same runs as in Fig.~\ref{fig:butter2}.}%
\label{fig:phase}
\end{minipage}
\end{center}
\end{figure}

The phase diagram of the horizontal components of the large-scale
field averaged over $0.2d<z<0.8d$ in Run~A2 are shown in
Fig.~\ref{fig:phase}(a). The streamwise and cross-stream field
components are in antiphase in this case.
This is expected because $S$ is negative \citep{Stix76}, so a positive
$B_x$ produces a negative $B_y$ when $S<0$.

\subsubsection{Summary}
\label{Summary31}
\begin{figure}
\begin{center}
\begin{minipage}{170mm}
\subfigure[]{
\resizebox*{8.5cm}{!}{\includegraphics{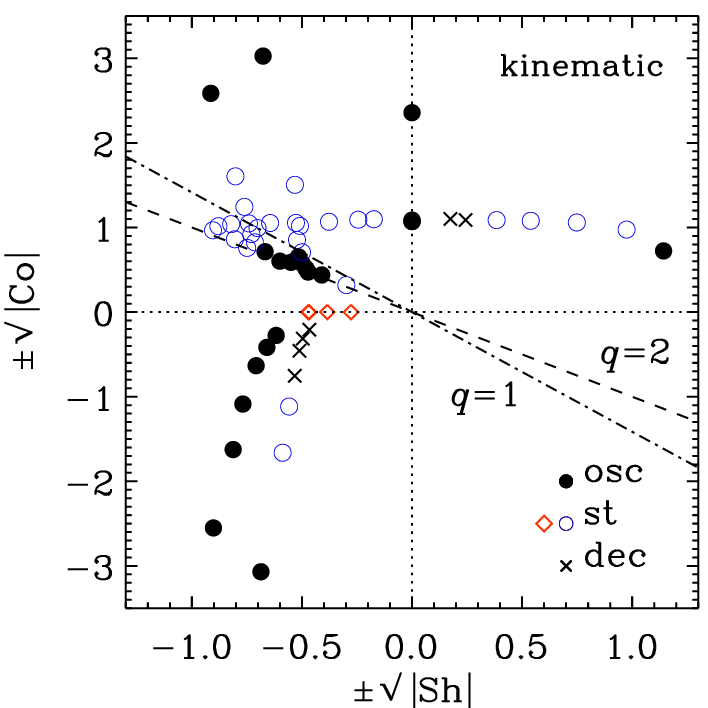}}}%
\subfigure[]{
\resizebox*{8.5cm}{!}{\includegraphics{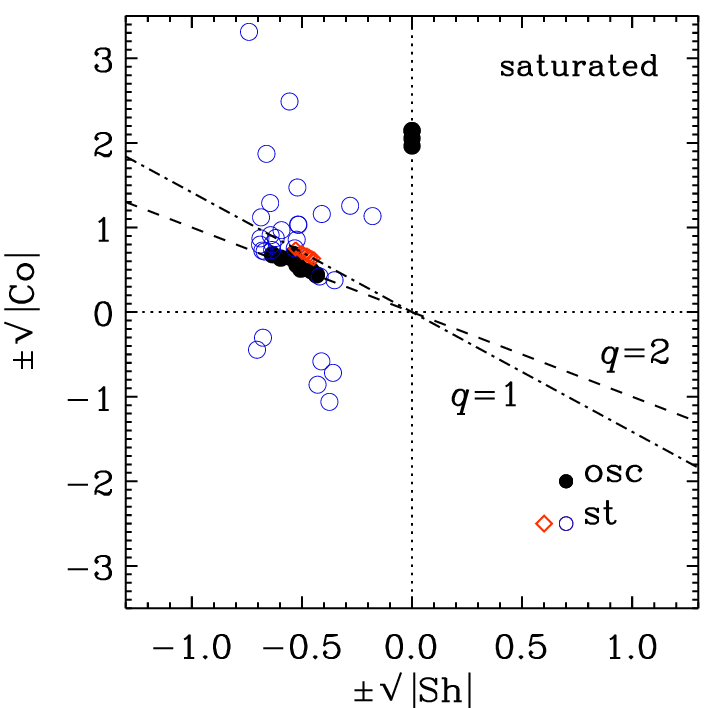}}}%
\caption{Dynamo mode in the kinematic regime (a) and in the
  saturated stage (b).
  Filled circles indicate oscillatory solutions and
  open circles (red and blue) quasi-steady ones.
  Red circles are data from \cite{KKB08}.
  Crosses indicate decaying solutions.
  Dashed and dash-dotted lines mark the positions of
  $q=1$ and 2, respectively}%
\label{fig:pmap_sat}
\end{minipage}
\end{center}
\end{figure}

The occurrence of oscillatory and quasi-steady solutions is conveniently discussed
in parameter space where these types of solutions are marked in a $\Sh$--$\Co$ diagram,
see in Fig.~\ref{fig:pmap_sat}(a) and (b) for the kinematic and saturated
regimes, respectively.
Both coordinate axes are stretched by taking the square root, i.e.,
we use $\pm\sqrt{|\Sh|}$, where upper and lower signs refer to the
sign of $\Sh$ (and likewise for $\Co$).
Oscillatory solutions mainly occur in two branches near $q=2$
for $-\Sh>0.2$ and near $\Sh=0$ for high enough $\Co$ 
(see Sect.~\ref{sect:setD}).
This is also true of the kinematic regime, but in that case there are additional
occurrences of oscillatory solutions for strong negative shear and
both signs of $\Co$; see Fig.~\ref{fig:pmap_sat}(a).
Results from spherical geometry suggest that the appearance of cyclic 
magnetic fields depends also on the magnetic Reynolds number 
\citep{BBBMT11}.

\subsection{Oscillatory $\alpha^2$ dynamos}\label{sect:setD}
In an earlier study we found the appearance of large-scale magnetic
fields in rigidly rotating convection \citep{KKB09b}. However, none of
the runs in that paper were run for much more than $10^3$ convective
turnover times. Although sign changes of the large-scale fields were
seen (see, Fig.~7 of KKB09), the time series were too short to
enable any firm conclusions regarding the possibly oscillatory nature
of the dynamo.

Furthermore, in similar rapidly rotating runs without magnetic fields,
the appearance of large-scale vortices has been discovered
\citep{C07,KMH11,MKH11}.
Here we use the hydrodynamical states of runs with large-scale
cyclones as initial conditions for our dynamo simulations. We find
that a large-scale dynamo is excited provided the magnetic
Reynolds number exceeds a certain critical value. Furthermore, as the magnetic
fields become dynamically important, the cyclones decay and are
absent in the non-linear stage. The large-scale magnetic field is
oscillatory in the two cases with different values of $\Co$ that we have
considered.
We conjecture that the dynamo is of $\alpha^2$ type in which case
  oscillatory solutions can be excited if the $\alpha$-effect has a
  suitable spatial profile \citep[e.g.][]{BS87,REO03,MTKB10}.
Figure~\ref{fig:butter2}(b) shows the horizontally averaged mean magnetic
fields from a rigidly rotating Run~D1 where an $\alpha^2$ dynamo is
excited.
In Run~D1 the large-scale fields are only functions of $z$ whereas in
the more rapidly rotating Run~D2 the large-scale fields depend also on
$x$ and $y$. Furthermore, the oscillatory nature of the solution is
then not so clear.
Figure~\ref{fig:phase}(b) shows the phase diagram of the horizontal
components of the large-scale field in Run~D1. There is a phase shift
of $\pi/2$.

The saturation level of the dynamo is sensitive to the magnetic
Reynolds number. Decreasing $\Rm$ from 66 to 39 by doubling the value
of $\eta$, decreases the saturation field strength by a factor of
three (Run~D1b). Another doubling of $\eta$ shuts the dynamo off
(Run~D1c).

Our standard setup in the present paper is to use perfect conductor
boundaries at the bottom and vertical field conditions at the
top. Changing the lower boundary also to vertical field conditions
produces no discernible difference in the solution (Run~D1d). However,
imposing perfect conductor conditions on both boundaries decreases the
saturation strength to less than a half of that in the standard setup and
decreases the fraction of the large-scale field (Run~D1e),
but the solutions remain oscillatory.
We have not, however, studied the $\Rm$-dependence of the saturation
field strength in this case, as was done in \cite{KKB10}.

\section{Conclusions}

We have presented results from simulations of turbulent magnetized convection
both with an imposed shear flow using
the shearing box approximation (Sets~A, B, E, and F) and in rigidly rotating
cases (Set~D). 
In accordance with previous results, we find the
generation of dynamically important large-scale magnetic fields. In
distinction to our earlier studies, we vary here the relative
importance of rotation and shear by covering the range
$q=-10\ldots1.99$ of the relative shear rate $q$. We find that for
$q=1$ the solutions are always stationary, which is in accordance with earlier
results. In Sets~A and B, where the shear is kept constant and
the rotational influence is varied, oscillatory solutions are found
for large $q$, i.e.\ slower rotation, and for $q\ll1$.
These trends are particularly clear in the kinematic regime. In the
saturated state, however, we often find that the dynamo switches from
oscillatory to stationary.
In Set~C, where $\varOmega_0={\rm const}$, only a single run
  shows oscillatory magnetic field in the saturated regime. Keeping
  $q$ fixed and varying $\Co$ and $\Sh$ shows that oscillatory
  solutions appear only in the rather narrow range $0.24<\Co<0.41$.
  Furthermore, when a perfect conductor boundary condition is adopted
  also at the top, the dynamo changes to a quasi-steady mode.
  Similar dependencies on boundary conditions can also be found in
  mean-field dynamos and are usually in agreement with corresponding DNS.
  These results
  suggest that a more thorough parameter study is needed and that the
  direct simulations need to be compared with mean-field models with
  the same parameters and turbulent transport coefficients from the
  test-field method. However, such a study is out of scope of the
  present paper.

It might not be easy to find general rules governing the transitions
from oscillatory to non-oscillatory behaviour in the parameter space
defined by shear and rotation rates in the kinematic and nonlinear regimes.
The simple rule that dynamos with shear oscillate while those without
shear do not, is only safe in homogeneous systems without boundaries.
For example,
in the rigidly rotating cases of Set~D, all the dynamo solutions are
found to be oscillatory.
In some cases, suitable spatial profiles of the resulting $\alpha$ effect
have been found to be responsible for oscillatory behaviour
\citep[e.g.][]{BS87,REO03,MTKB10}.
Large-scale vortices, present in the
hydrodynamic state, are no longer found in the saturated state of the
dynamo. Usage of a perfect conductor boundary condition instead of a
vertical field condition, allowing for magnetic helicity fluxes, is
found to decrease both the total saturation field strength and the
strength of the large-scale field with respect to the total magnetic
field.
This might be a consequence of what is known as catastrophic
(or $\Rm$-dependent) quenching which cannot easily be alleviated in a
closed domain, but it might also be a consequence of a delayed onset
of dynamo action, which is explained by linear theory.
In fact, recent work on catastrophic quenching has shown that
resistive effects tend to dominate over effects resulting
from magnetic helicity fluxes for magnetic Reynolds numbers below
a value of around $10^3$ or even $10^4$ \citep{CHBM11}.
This makes an explicit demonstration of alleviated catastrophic
quenching hard at the $\Rm$ values available to date.

\section*{Acknowledgements}

We thank the three anonymous referees for making useful suggestions.
The computations have been carried out using
the facilities hosted by the CSC  -- IT Center for Science in Espoo, Finland,
who are financed by the Finnish ministry of education.
This work was supported in part by
the European Research Council under the AstroDyn Research Project No.\ 227952,
the Swedish Research Council Grant No.\ 621-2007-4064, and the Academy
of Finland grants 136189, 140970 (PJK) and 218159, 141017 (MJM).


\newcommand{\yppN}[4]{, #4, {\em Phys.\ Plasmas }#1, {\bf #2}, #3.}
\newcommand{\yapjN}[4]{, #4, {\em Astrophys.\ J.\ }#1, {\bf #2}, #3.}
\newcommand{\yapj}[5]{, #5, {\em Astrophys.\ J.\ }#1, {\bf #2}, #3-#4.}
\newcommand{\yan}[5]{, #5, {\em Astron.\ Nachr.\ }#1, {\bf #2}, #3-#4.}
\newcommand{\yana}[5]{, #5, {\em Astron.\ Astrophys.\ }#1, {\bf #2}, #3-#4.}
\newcommand{\ymn}[5]{, #5, {\em Monthly Notices Roy.\ Astron.\ Soc.\ }#1, {\bf #2}, #3-#4.}


\label{lastpage}

\end{document}